# From 5G to 6G: Revolutionizing Satellite Networks through TRANTOR Foundation

*Pol Henarejos[1]\*, Xavier Artiga[1], Miguel A. Vázquez[1], Màrius Caus[1], Musbah Shaat[1], Joan Bas[1], Lluís Blanco[1], and Ana I. Pérez-Neira[1]*

[1]*Space and Resilient Communication Systems Research Unit (SRCOM), Centre Tecnològic de Telecomunicacions de Catalunya (CTTC), Av. Carl Friedrich Gauss 7, Castelldefels, Spain*

\*pol.henarejos@cttc.es

## Abstract

5G technology will drastically change the way satellite internet providers deliver services by offering higher data speeds, massive network capacity, reduced latency, improved reliability and increased availability. A standardised 5G ecosystem will enable adapting 5G to satellite needs. The EU-funded TRANTOR project will seek to develop novel and secure satellite network management solutions that allow scaling up heterogeneous satellite traffic demands and capacities in a cost-effective and highly dynamic way. Researchers also target the development of flexible 6G non-terrestrial access architectures. The focus will be on the design of a multi-orbit and multi-band antenna for satellite user equipment (UE), as well as the development of gNodeB (gNB) and UE 5G non-terrestrial network equipment to support multi-connectivity.

## 1 Background

Fifth generation non-terrestrial network (5G NTN) began standardization activities in 2018 with great technological interest from the terrestrial manufacturers, allowing a satellite-terrestrial radio access network convergence. Since the introduction of the 5G NTN standard peculiarities, new and more advanced generations of radio access and network management techniques have emerged for its application into diverse vertical sectors: maritime, automotive, etc. Despite this optimistic market picture, the crude reality is that satellite operators are only slowly adopting basic 5G deployments. Facing this reality, it can be claimed that there is no clear path to introduce 5G in satellite networks. Several barriers are currently blocking the adoption of this technology. The first 5G adoption barrier is related to satellite operator culture. The workforce typically consists of teleport engineers, who deal with monolithic radio frequency management issues. 5G NTN will make radio access engineers the main force: they will deal with a more automated network, avoiding ticketing mechanisms for incorporating new clients to the network system using software configuration tools. While it is clear that network automation is needed for network operators to fully benefit from 5G NTN adoption, to enable such automation the key is to define use cases and workflows with standard protocols that operate in the greenfield and brownfield deployments. During the last few years, we have observed the confluence of flexible payload missions and 5G networks, with flexible satellite adoption is slowly evolving but without a clear adoption path. Flexible payloads and 5G are two flips of the same coin, and the applicability of automated decisions in this context is uncertain when dealing with challenges in network management, user equipment, security, optimization, and scalability.

As 5G NTN networks are deployed, 3GPP release 18 and pre-6G NTN networking will require highly dynamic radio access management. Automated dynamicity is essential in these scenarios, in which humans will not be able to manage and operate these networks. 5G-Advance (5G-A) must provide the capabilities to fulfil these requirements. For example, satellite beam bandwidth control at the Network Operator Center (NOC) is not efficient enough. A massive number of clients with heterogeneous service level agreements, cannot be consistently handled with current teleport controller solutions. These solutions consist of a web-based human-centered systems that work on event-based (outage alerts). Some limitations to this current software architecture have been raised, and ground segment manufacturers are slowly looking at possible solutions, by completely redesigning NOC controllers. But only considering inter-operability mechanisms for data flow management is not enough as there is also a clear need for radio access control in support for 5G-A scenarios. A significant challenge is the joint addressing the radio access and the satellite operations control in a unified way. Satellite resources and radio access represent a continuum within all potential network topologies, where user equipments (UE) tend to be constraint and spread geographically. This also leads to the need for proper integration of 5G Core Network (CN) procedures in NOCs to be applied to pre-6G networks.

NTN 5G-A and future pre-6G networks must be understood as 3D multi-layered networks constituted by multiple satellites deployed in multiple orbits (GEO, MEO, LEO) employing the multiple frequency bands available by regulation. Multiorbital satellite networks require the introduction of UE components that are able to point to the intended satellite while facing challenging propagation impairments. This leads to a novel UE 5G NTN re-design with a user-centric design to reduce terminal costs and enhance the throughput and resilience. Several challenges show up in here ranging from the antenna aperture cost-effective solutions for dealing with a diverse set of link budgets and the beam-pointing mechanisms which shall both provide a short initial access delay and seamless track of the moving satellites. Missions employing multi-band payloads offer and additional flexibility for adverse propagation environments where lower frequency communications is more advantageous. In all context, end-to-end satellite 5G-A NTN operation shall be merged with NOCs messaging and telemetry controls together with the 5G CN.

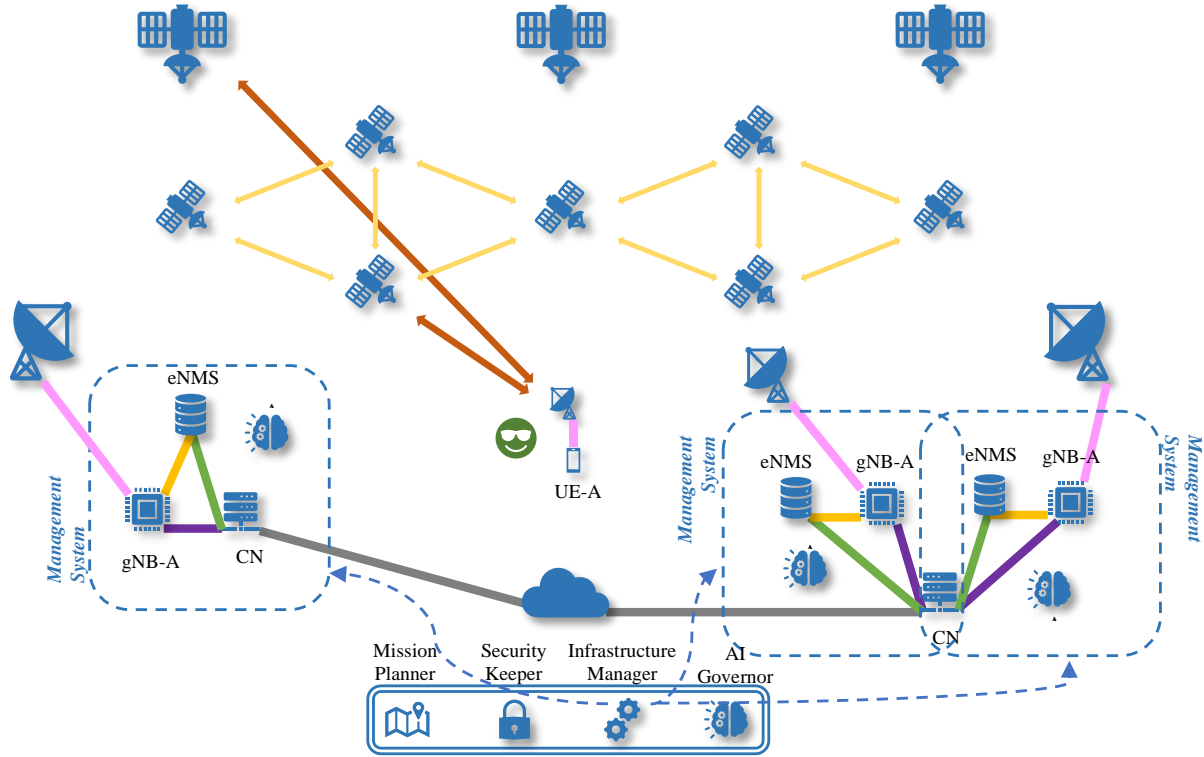

*Figure 1. TRANTOR overall concept.*

## 2   Implementation and Methodology

Although the definition of 6G networks is still fully open, three of their main pillars can be already envisaged. First, 6G networks will become truly UE-centric communication systems. This means that the UE will not adapt to the network needs, but that the network will seamlessly manage all its available resources to provide the UE required throughput and latency, whichever they are, whenever the UE asks for them and wherever the UE is located. Second, the satellite segment in 6G networks will be constituted by an overcrowded 3D-layered network with multiple satellites deployed in multiple orbits at different altitudes, using multiple frequency bands. Third, an automated and data driven network management system will be required to assign dynamically and efficiently the network resources to the served users.

This vision on 6G networks directly defines the TRANTOR (5G+ evoluTion to mutioRbitAl multibaNd neTwORks) concept, which is sketched in Figure 1. The main idea is to develop and validate a unique/unified network management system across the 3D multi-layered satellite network. The resulting homogeneous network will be able to manage multiple efficient radio connections, gathering the synergies of multiple satellite, multiple orbit and multiple frequency communications, providing transmission diversity and ensuring maximum availability and the required throughput to every user. Indeed, TRANTOR focuses on the provision of efficient broadband services to advanced satellite UE terminals, using Ka and Ku bands. The rationale behind these advanced UE terminals is that, providing them new multiband and multilink capabilities, they will have access to an unprecedented number of resources of the 3D network for satisfying their instantaneous needs.

Moreover, TRANTOR is conceived as an extension of 5G NTN towards 6G. In this way, the 5G NR architecture, procedures and waveforms are fully adopted and will be extended not only to support basic NTN operation, but also to support advanced multi-band and multi-link use cases as well as a truly integrated and automated network management, which must be considered as a 5G-A or pre-6G solution. Although TRANTOR is only devoted to the satellite segment, the adoption of 5G framework ensures the convergence of space and terrestrial networks and services.

TRANTOR relies on the following key enabling components, which will be developed during the project lifetime and constitute the main project contributions:

**Advanced UE satellite terminal (UE-A):** It consists of 1) a 5G-A modem including extensions to support basic GEO and NGEO operation as well as multi-link and multi-band capabilities; 2) an advanced antenna aperture with electronical steering and multi-beam, multi-band and multiorbital capabilities; 3) an evolved antenna control unit for accurate antenna pointing based only on 5G NR waveform and 5G NR beam management procedures.

**Advanced gNB (gNB-A):** Extending basic terrestrial gNB functionalities to support GEO and LEO connectivity, considering for instance the adaptation of numerology for fast acquisition and tracking, the adaption of synchronisation procedures, the impact of long round trip times, considerations on CU/DU splitting, etc.

**Evolved satellite Network Manager System (eNMS) and Integrated management system:** Traditional satellite NMS needs to be evolved in two directions: 1) to support highly dynamic automated and data driven resource management; and 2) to be fully integrated in the 3GPP management system. To this end, clear interfaces between NMS, 5G core and gNB need to be developed and an efficient functional split between the three entities needs to be developed.

**Infrastructure Manager Entity (IME):** The 3D multi-layer network is constituted in fact by multiple satellite systems (i.e., GEO, MEO or LEO) each of them with its own NMS to control its own resources. TRANTOR will thus develop and IME to manage efficiently all the available resources across the multiple systems. Although plotted as a centralized entity in *Figure 1*, the IME could be also thought of a set of functionalities split between a centralized entity and the distributed evolved NMS.

**Artificial Intelligence (AI) governance modules:** The large number of users and resources available across the multiple satellite systems requires and automated data-assisted framework to perform a proper resource assignment. In this sense, TRANTOR will develop AI modules to assist traffic and resource management both at IME level and local 3GPP management system level (i.e., CN, gNB and NMS integrated management).

**Mission Planner:** It consists of a system level simulation platform that will allow a priori evaluating and predicting the performance of using resources belonging to multiple satellite systems, as well as optimizing AI governance modules. It will become not only a research tool but also a required tool for operators planning their future system infrastructures.

**Security Keeper:** It consist of a Thread risk assessment, performing end-to-end cyber security risk assessment and a Dynamic risk mitigator putting in practice thread countermeasures. It will be focused to new thread types arising from the integrated management and operation of multiple satellite systems.

## 3 Experimental Approach

TRANTOR proposal has been built with a clear experimental validation orientation. In this way, a two phases demonstration framework is proposed. The first phase, to be performed at project mid-term, will be devoted to individual systems connectivity to ensure that the performance and requirements are fulfilled when working with separated systems. End-to-end 5G NR transmissions over GEO, a drone-emulated LEO and OBP CU/DU splitting capabilities will be tested independently. On the second phase, to be performed at project end, systems will be integrated to introduce the communication convergence across networks, providing a unique management. In this step, systems will be tested jointly enabling multi-band, multi-GEO and multiorbital connectivity, so that diversity, high performance and integrated management can be validated and evaluated. Targeting satellite network evolution towards 6G would require full consideration of LEO mega-constellations with enhanced on-board capabilities such as traffic routing through ISL or in-orbit edge computing. However, the current lack of an EU operated LEO network does not permit in-orbit experimentation on these topics, so the TRANTOR approach will be to address them with a longer time frame target, thus reaching lower TRL for the LEO involvement.

The TRANTOR approach represents a radical yet credible step beyond current state-of-the-art of satellite systems. Current satellite systems are monolithically and independently operated, each of them counting with its own NMS managing all its own available resources. Moreover, the NMS operation is not fully automated, often requiring the participation of operation engineers thus not supporting a highly dynamic adaptability to user requirements. Although the jump to a unified management network across diverse satellite systems is huge, several recent activities indicate that the satellite community bets for this solution. In particular, there have been already several trials involving multiple operators to provide multiorbital connectivity [1][2][3], and remarkably Inmarsat announced Orchestra [4], a multi-satellite multiorbital system manager. Besides, several operators and ground segment providers are involved in research activities to develop AI based resource management solutions to automatize part of their NMS functions, especially in the context of flexible payloads, such as EU funded project ATRIA[5].

Note however, that none of the multiorbital trials mentioned before included 5G NR transmissions. Indeed, although the satellite segment demonstrated a significant interest on the adoption of 5G, only a few end-to-end 5G NR over satellite tests have been reported. Therefore, the full adoption of 5G NR in the satellite segment is still far. The standardization activities at 3GPP have already identified the minimum set of necessary features enabling NR support for NTN in Release 16 and based on this an ongoing work item in Release 17 is dealing with multiple aspects from, physical layer to architecture, protocols, etc., but their standardization is not even closed. Moreover, the advanced functionalities covered by TRANTOR such as multi-band and multi-link (i.e., multiTRP using 3GPP terminology) have not been considered. The integration of the NMS on the 3GPP management system is in a similar state, the 3GPP has identified the need of integration but the functional split between the satellite NMS, 5G core and gNB have been not solved.

On the UE side, and besides the adoption an extension of 5G NR framework, the development of an electronic smart antenna array able to cope with both GEO and NGEO satellites through hybrid beamforming, to work with multiple satellites simultaneously, in different frequency bands, and including real-time satellite tracking capabilities even in mobility scenarios, represents a significant step beyond current solutions. Although strong efforts have been performed in the recent years in the design of planar electronically scanned arrays for satellite-on-the-move terminals, current solutions are limited to single band single beam capabilities, with pointing mechanisms not directly relaying on 5G NR waveforms or beam management procedures.

Summarizing, TRANTOR outcomes will provide a clear advance with respect to SOTA in terms of 5G adoption in satellite systems, 5G and satellite network management integration, resource management automation across multiple satellite systems, support to multi-band multi-link connectivity frameworks, advanced UE functionalities and secure operation across multiple satellite systems.

*3.1 Demonstration 1 (Phase 1a): End-to-End single band connectivity with a single GEO satellite*

**Rationale**: Baseline scenario demonstrating the basic support to end-to-end 5G NR connectivity over a GEO satellite.

**Description**: It is composed by an end-to-end scenario, where a UE is able to communicate with internet via a GEO satellite using 5G-NR waveform. It will validate the extended capabilities provided to the gNB-A and UE-A for single link 5G NTN support. They include the modem baseband modifications as well as a preliminar functional UE-A antenna prototype.

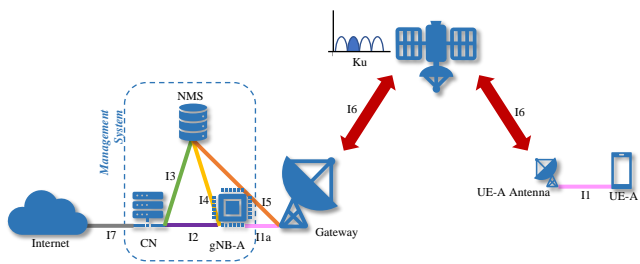

*Figure 2. Demonstration 1 scheme.*

*3.2 Demonstration 2 (Phase 1a): End-to-End single band connectivity with a single drone-emulated LEO satellite*

**Rationale**: This scenario is similar to Demonstration 1, but with a LEO satellite instead of GEO.

**Description**: Besides the capabilities already validated in the GEO case, this demonstration will permit demonstrating UE-A capabilities to track a LEO satellite using 5G NR waveform and beam management procedures as well as to compensate the high Doppler shift produced by the low-high satellite orbit. Therefore, two major challenges are present in this scenario: pointing capabilities and Doppler shift compensation, even before the initial handshake to the network occurs. As stated before, a drone-Emulated LEO framework will be adopted to test connectivity and tracking with the UE-A. Therefore, no NMS will be used and the gateway will just perform upconversion and transmission to the drone.

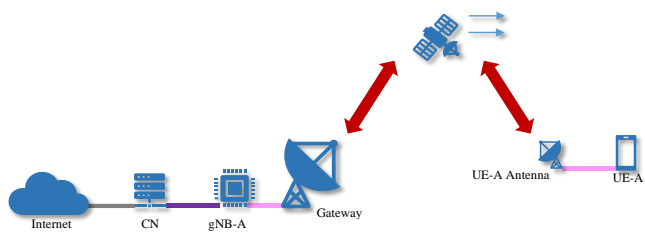

*Figure 3. Demonstration 2 scheme.*

*3.3 Demonstration 3 (Phase 1b): End-to-End single band connectivity with CU/DU split with OBP satellite*

**Rationale**: This scenario is similar to Demonstration 1 but exploiting OBP capabilities.

**Description**: In this demonstration, the concept of CU/DU splitting, with part of the signal processing performed in ground and the rest on board on the satellite, is showcased. This demonstration is a mid-step between the phase 1a and 2, which culminates the progresses made in the next generation satellites tasks. This scenario presents two major challenges: find the optimal splitting between CU and DU amongst all possibilities and the implementation of part of the signal processing blocks from the ground to the space, which is addressed with hardware and software programming rather than a fully software defined radio approach.

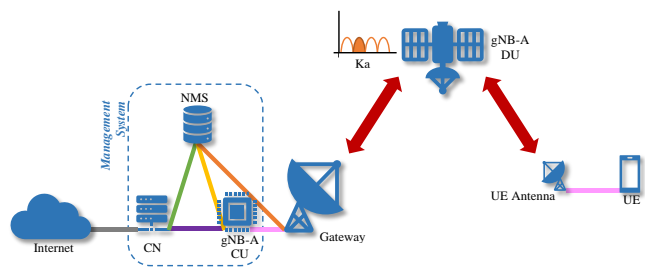

*Figure 4. Demonstration 3 scheme.*

*3.4 Demonstration 4 (Phase 2): Multi-band transmission from a single GEO satellite*

**Rationale**: This scenario is an extended multiend case, where the UE has Advanced capabilities to receive in multiple bands, specifically Ku and Ka.

**Description**: In this scenario, the gNB-A is configured with multiple cells with central carriers placed in Ku and Ka bands in the same satellite. Therefore, the UE-A is able to either hand over through Ku or Ka bands depending on channel conditions or for efficient resource sharing or receive in Ku and Ka bands simultaneously for increased capacity. The UE-A antenna has fully operational multiband capabilities and the eNMS is integrated in the management system, which performs dynamic management of the multi-band resources, as well as traffic management in the case simultaneous multi-band links.

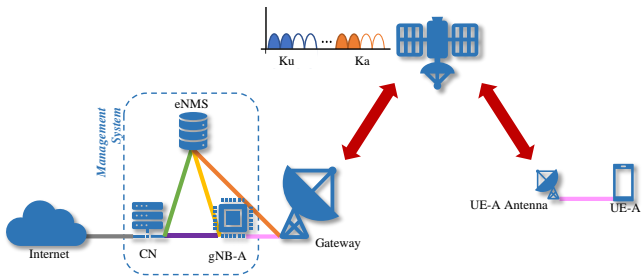

*Figure 5. Demonstration 4 scheme.*

*3.5 Demonstration 5 (Phase 2): Multi-satellite, multi-band transmission using two GEO satellites*

**Rationale**: this demonstration extends the previous scenario by considering that the two bands are transmitted from different GEO satellites, thus the UE-A has Advanced multi-satellite (i.e., multiTRP capabilities).

**Description**: this scenario depicts a multiend scenario with two GEO satellites. The UE-A is able to point to different GEO satellites autonomously, either performing hand over between them or supporting simultaneous links. The eNMS is able to manage dynamically the frequency resources across the two satellites, according to channel conditions or other resource sharing needs. In addition, the integrated Management system and more specific the traffic manager needs to deal with the different propagation delays for each path.

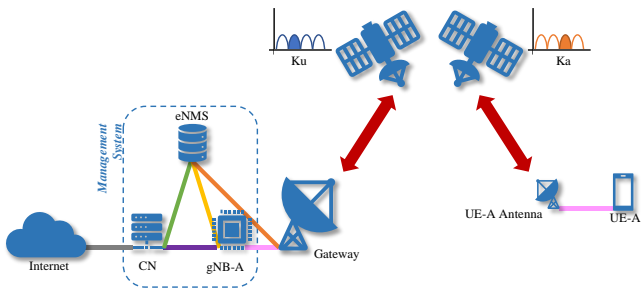

*Figure 6. Demonstration 5 scheme.*

*3.6 Demonstration 6 (Phase 2): Multiorbital, multi-band transmission using a GEO and a drone-emulated LEO satellites*

**Rationale**: this scenario puts together all previous scenarios considering a multi-band transmission from a LEO and a GEO satellite.

**Description**: On top of the TRANTOR capabilities already validated in previous demonstrations, this framework requires and permits validating the efficient use of the UE-antenna aperture resources for multiorbital connectivity. Besides, it requires the basic functionality of an IME, managing resources across two independent satellite systems. In this case, the frequency resource and traffic management will not only consider the channel conditions or the efficient use of resources but also the large difference between the propagation path across the LEO and GEO orbits and the latency requirements of the user streams (i.e., low latency packets must be transmitted over LEO network).

In this demonstration framework, involving multiple satellite systems and the presence of an IME on top of them, we will also validate the main functionalities of TRANTOR security keeper, to demonstrate a secure management across the heterogeneous network. In addition, a comprehensive TRA campaign to assess and validate network security resilience will be performed.

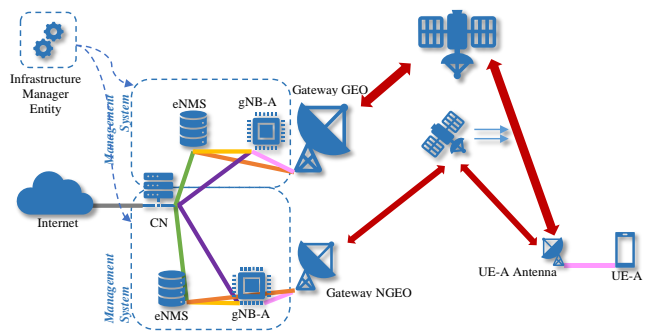

*Figure 7. Demonstration 6 scheme.*

## 4 Conclusions

In conclusion, the implementation of 5G and its evolution towards 6G in satellite networks presents significant challenges but also promising opportunities. Satellite operators face barriers related to culture, network automation, and integration of radio access and satellite operations. However, the TRANTOR concept offers a solution with a unified network management system that can efficiently handle the complexities of multi-layered satellite networks, providing advanced broadband services and transmission diversity. By developing advanced UE satellite terminals, evolved gNBs, AI governance modules, and other key components, TRANTOR aims to converge space and terrestrial networks, paving the way for the seamless integration of 6G and beyond in satellite systems.

## 5 Acknowledgements

This work has been supported by the project TRANTOR which has received funding from the European Union's Horizon Europe research and innovation program under grant agreement No. 101081983.